\documentclass[twocolumn,floatfix,prx,aps,showpacs]{revtex4-1}
\usepackage{graphicx,amsmath,amssymb,color,xcolor}
\usepackage{nicefrac}
\usepackage[titletoc,title]{appendix}

\newcommand{\be}{\begin{equation}}
\newcommand{\ee}{\end{equation}}

\newcommand{\ba}{\begin{eqnarray}}
\newcommand{\ea}{\end{eqnarray}}

\begin{document}

\title{Fractional Quantum Hall Effect at $\nu=2+6/13$: The Parton Paradigm for the Second Landau Level}
\author{Ajit C. Balram$^{\dagger 1}$, Sutirtha Mukherjee$^{\dagger 2}$, Kwon Park$^{2,3}$, Maissam Barkeshli$^{4}$, Mark S. Rudner$^{1}$ and J. K. Jain$^{5}$}
\affiliation{$^{1}$Niels Bohr International Academy and the Center for Quantum Devices, Niels Bohr Institute, University of Copenhagen, 2100 Copenhagen, Denmark}
\affiliation{$^{2}$Quantum Universe Center, Korea Institute for Advanced Study, Seoul 02455, Korea}
\affiliation{$^{3}$School of Physics, Korea Institute for Advanced Study, Seoul 02455, Korea}
\affiliation{$^{4}$Condensed Matter Theory Center and Joint Quantum Institute, Department of Physics, University of Maryland, College Park, Maryland 20472 USA}
\affiliation{$^{5}$Department of Physics, 104 Davey Lab, Pennsylvania State University, University Park, Pennsylvania 16802, USA}

\date{\today}

\begin{abstract} 
The unexpected appearance of a fractional quantum Hall effect (FQHE) plateau at $\nu=2+6/13$~ [Kumar \emph{et al.}, Phys. Rev. Lett. {\bf 105}, 246808 (2010)] offers a clue into the physical mechanism of the FQHE in the second Landau level (SLL). Here we propose a ``$\bar{3}\bar{2}111$'' parton wave function, which is topologically distinct from the 6/13 state in the lowest Landau level. We demonstrate the $\bar{3}\bar{2}111$ state to be a good candidate for the $\nu=2+6/13$ FQHE, and make predictions for experimentally measurable properties that can reveal the nature of this state. Furthermore, we propose that the ``$\bar{n}\bar{2}111$'' family of parton states naturally describes many observed SLL FQHE plateaus. 
\end{abstract}
\maketitle

Ever since its discovery more than three decades ago, the fractional quantum Hall effect (FQHE)~\cite{Tsui82} has provided a fertile playground to study quantum many-body phenomena. Intriguingly, fundamentally distinct physics underlies the FQHE in different Landau levels (LLs). For the lowest LL (LLL), a unified understanding of the FQHE has been developed in terms of composite fermions (CFs): the prominent FQHE at filling factor $\nu=n/(2pn\pm 1)$, where $n$ and $p$ are positive integers, arises as the $\nu^*=n$ integer quantum Hall effect (IQHE) of composite fermions~\cite{Jain89}, while weaker plateaus at fractions such as $\nu = 4/11$ arise as FQHE states of interacting composite fermions~\cite{Park00b, Pan03,Wojs04,Mukherjee14}. In the half filled LLL, $\nu = 1/2$, a {\it compressible} Halperin-Lee-Read Fermi sea of composite fermions~\cite{Halperin93} is realized. In striking contrast, the half filled second LL (SLL) hosts an {\it incompressible} FQH state at $\nu = 2 + 1/2$~\cite{Willett87}. Numerical studies have supported the notion that this state is a paired state of composite fermions, described by the Moore-Read Pfaffian wave function~\cite{Moore91} or its particle-hole conjugate, the anti-Pfaffian~\cite{Levin07,Lee07}. Recent thermal Hall measurements~\cite{Banerjee17b} appear to be inconsistent with both of these candidate states; a number of scenarios to explain the measurements are currently being debated~\cite{Alicea18}. 

The surprising observation of a FQHE at $\nu=2+6/13$ by Kumar \emph{et al.}~\cite{Kumar10} further underscores the difference between the LLL and the SLL. In addition to the fact that the $2+1/2$ state is not a CF Fermi sea, experiments do {\it not} show conclusive evidence for FQHE at $\nu = 2 + 3/7$, $2 + 4/9$, and $2 + 5/11$~\cite{Shingla18}; the FQHE at $\nu = 2 + 2/5$~\cite{Xia04,Choi08,Pan08,Kumar10} is furthermore believed to be parafermionic and not CF-like~\cite{Read99,Rezayi09,Sreejith13,Zhu15,Mong15,Pakrouski16}, and even the nature of the state at $2 + 1/3$ is under debate~\cite{Ambrumenil88,Balram13b,Johri14,Peterson15,Jeong16}.
Based on all of these observations, we conclude that a CF-based LLL-like description of the $\nu=2+6/13$ state is highly unlikely. This motivates us to search for a unifying principle for the FQHE in the SLL. In this work we propose and analyze the $\nu = 2 + 6/13$ FQHE in terms of a ``parton state" that is topologically distinct from its LLL counterpart. We demonstrate that the parton wave function has lower energy than the LLL CF $\nu = 6/13$ state for the ideal SLL Coulomb interaction (neglecting disorder, LL mixing and finite width), and also a reasonably high overlap with the ground state obtained via exact diagonalization. Taking these results together with the recent demonstration of a related parton wave function for the ground state at $\nu = 2 + 1/2$, we propose a sequence of parton wave functions that naturally captures many prominent fractions observed in the SLL. We provide several experimentally testable predictions that follow from our proposal.

\begin{figure*}[t]
\begin{center}
\includegraphics[width=1\textwidth,height=0.24\textwidth]{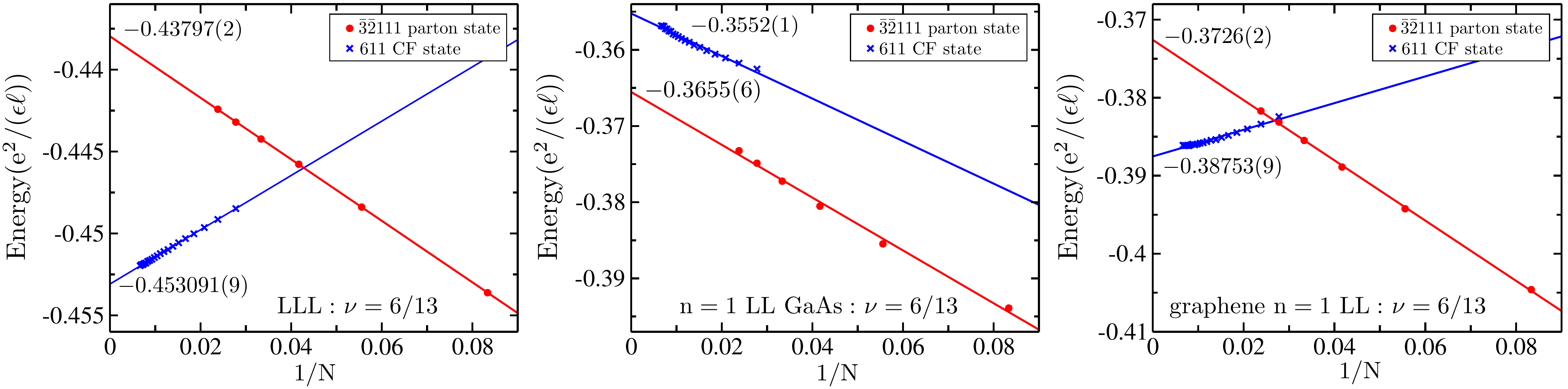} 
\caption{(color online) Thermodynamic extrapolations of the energies (per particle) for the CF state (blue crosses) and the $\bar{3}\bar{2}111$ parton state (red dots). The left-hand panel show energies for $\nu=6/13$ in the LLL, the middle panel for $\nu=2+6/13$ in the second LL, and the right-hand panel for $6/13$ in the $n=1$ LL of monolayer graphene (for $n=0$ LL, graphene results are identical to those in the left-hand panel for the $n=0$ GaAs LL). The energies include the electron-background and background-background interaction, and are quoted in units of $e^2/(\epsilon\ell)$. The LLL Coulomb energy for the CF state has been reproduced from Ref.~\cite{Balram17}. 
 }
\label{fig:extrapolations_energies_6_13}
\end{center}
\end{figure*}

Using the parton theory~\cite{Jain89b,Blok90,Blok90b,Jain90,Wen91,Wen92b}, we construct new FQHE states by decomposing each electron into fictitious particles called partons, placing each parton species into an IQHE state, and then fusing the partons back into physical electrons. The $m$-parton ``$n_{1}n_{2}\cdots n_m$'' wave function of $N$ electrons at filling factor $\nu$ is given by
\be
\label{eq:general_parton}\Psi^{\{n_\lambda\}}_{\nu}= \mathcal{P}_{\rm LLL} \prod_{\lambda=1}^m \Phi_{n_\lambda}(\{z_j\}),
\ee
where $z_j = x_j - i y_j$ describes the two-dimensional coordinates of electron $j$, with $1 \le j \le N$, and $\mathcal{P}_{\rm LLL}$ denotes projection into the LLL. Here $\Phi_{n_\lambda}$ is the IQHE wave function of $N$ electrons filling $n_{\lambda}$ LLs. We allow $n_\lambda < 0$, referring to IQHE states in a negative magnetic field: $\Phi_{\bar{n}}=[\Phi_n]^*$, where $\bar{n}$ denotes a negative value. To ensure that each parton species occupies the same physical area when exposed to the external magnetic field $B$, we must take their charges to be $q_\lambda=-e\nu/n_\lambda$, with $\sum_\lambda q_\lambda=-e$, where $-e$ is the electron charge. The physical filling factor of the state $\Psi^{\{n_\lambda\}}_\nu$ in Eq.~(\ref{eq:general_parton}) is then given by $\nu=( \sum_\lambda n_\lambda^{-1} )^{-1}$. The parton states of the types $n11\cdots$ and $\bar{n}11\cdots$ correspond to CF states~\cite{Jain89} with Abelian excitations. More general states of the form in Eq.~(\ref{eq:general_parton}) can also accommodate non-Abelian excitations~\cite{Wen91}.

Our motivation for considering a parton state to describe the $\nu=2+6/13$ FQHE derives from the recent application of the parton construction to the $\nu = 5/2$ FQHE in Ref.~\cite{Balram18}. There, the parton state $\bar{2}\bar{2}111$ was shown to exhibit a substantial overlap with the SLL Coulomb ground state obtained by numerical exact diagonalization, as well as the anti-Pfaffian wave function~\cite{Levin07,Lee07}. Further arguments showed that the $\bar{2}\bar{2}111$ parton state and the anti-Pfaffian state describe the same phase. These results, together with the experimental observations described above, lead us to consider parton states of the form~\cite{footnote:two_thirds_variants}:
\be
\label{eq:parton_family} \Psi^{\bar{n}\bar{2}111}_{\nu=2n/(5n-2)}=\mathcal{P}_{\rm LLL}  \Phi_{\bar{n}}\Phi_{\bar{2}}\Phi_1^3.
\ee
The choice $n=2$ produces the state at $\nu=1/2$ considered in Ref.~\cite{Balram18}. We shall investigate the $n=3$ state of this sequence, which occurs at $\nu = 6/13$. This $6/13$ parton state is also applicable, through particle-hole conjugation, to the recently observed FQHE at $\nu=7/13$ in the $n=1$ LL of bilayer graphene~\cite{Zibrov16}.

We note here that Levin and Halperin~\cite{Levin09a} proposed to obtain a FQHE at $\nu=2+6/13$ as the first daughter in a hierarchy emanating from the anti-Pfaffian. Since no simple wave function follows from the Levin-Halperin construction it has not figured prominently in numerical studies of the SLL FQHE. Interestingly, below we show that the Levin-Halperin state and the  $\bar{3}\bar{2}111$ parton state are possibly topologically equivalent.

An advantage of the parton construction is that the wave function in many cases can be evaluated for very large systems, well beyond the sizes accessible to exact diagonalization~\cite{Balram18}. This efficient evaluation is possible because we can project these states into the LLL as
\be
\Psi^{\bar{n}\bar{2}111}_{\nu=2n/(5n-2)}\sim {[\mathcal{P}_{\rm LLL}\Phi_{\bar{n}}\Phi_1^2] 
[\mathcal{P}_{\rm LLL}\Phi_{\bar{2}}\Phi_1^2]\over \Phi_1}=
\frac{\Psi^{\rm CF}_{n/(2n-1)}\Psi^{\rm CF}_{2/3}}{\Phi_{1}}.
\label{eq:parton_barnbar2111}
\ee
The $\sim$ indicates that the wave function on the the right-hand side of Eq. (\ref{eq:parton_barnbar2111}) differs slightly from the definition in Eq. (\ref{eq:parton_family}) in how the projection to the LLL is implemented. We expect such details of the projection to have only a minor effect on the state; in particular, we expect that the universality class of the state should not be affected~\cite{Balram16b}.
The CF states $\Psi^{\rm CF}_{n/(2n-1)}$ can be evaluated for hundreds of electrons using the Jain-Kamilla projection~\cite{Kamilla96,Moller05,Jain07,Davenport12}.

Throughout this work we employ the spherical geometry~\cite{Haldane83} in which $N$ electrons move on the surface of a sphere, with a radial magnetic field emanating from a monopole of strength $2Q (h/e)$ at the sphere's center.  Incompressible quantum Hall states occur at flux values $2Q=\nu^{-1}N-\mathcal{S}$, where $\mathcal{S}$ is a topological number called the shift~\cite{Wen92}. These states are uniform on the sphere and thus have total orbital angular momentum $L=0$. The parton states $\Psi^{\bar{n}\bar{2}111}_{\nu=2n/(5n-2)}$ occur at $2Q = N(5n-2)/(2n) - (1-n)$; i.e., their shifts are given by $\mathcal{S}=1-n$. In particular, the $\bar{3}\bar{2}111$ parton state has a shift of $\mathcal{S}=-2$, distinct from the shift of $\mathcal{S}=8$ for the $611$ CF state that also occurs at $\nu=6/13$.

We first ask if the $\bar{3}\bar{2}111$ parton state is a plausible candidate to describe the $\nu = 2 + 6/13$ FQHE. To this end, we begin by comparing the parton and the CF states in the LLL and the SLL (see Fig.~\ref{fig:extrapolations_energies_6_13}). In our calculations, all states are written for the LLL; the SLL is simulated by using an effective interaction that has the same pseudopotentials in the LLL as the Coulomb interaction in the SLL. Here we use the effective interaction described in Ref.~\cite{Shi08}. We find that the $611$ CF state has a lower energy in the LLL, as expected, while the $\bar{3}\bar{2}111$ parton state has lower energy in the SLL. For completeness we have also investigated the competition between these two states in the $n=1$ LL of monolayer graphene. We find the 611 CF state to be favored here, consistent with the observation that FQHE states in the $n=1$ LL of monolayer graphene conform to the CF paradigm~\cite{Amet15,Balram15c,Zeng18}.

We next ask how accurately the parton state represents the ground state found from exact diagonalization. For this purpose we shall use the SLL Coulomb pseudopotentials of the disk geometry, which slightly differ from those of the spherical geometry but are known to give better thermodynamic extrapolation. For $N = 12$ particles, the overlap of the parton wave function with the numerically exact ground state is $0.7536(9)$. The ground state energy per particle of the parton state for the effective interaction is $-0.39390(6)$ while the exact energy is $-0.39689$, both in units of $e^2/(\epsilon \ell)$, where $\epsilon$ is the dielectric constant of the host material and $\ell=\sqrt{\hbar/(eB)}$ is the magnetic length. (This energy has been corrected for the finite size deviation of the density in the spherical geometry, and includes electron-background and background-background interactions.) The agreement by itself is not conclusive, but provides comparable evidence to that obtained from overlaps for the Pfaffian~\cite{Morf98,Scarola02,Pakrouski15} and Laughlin~\cite{Ambrumenil88,Balram13b} states at filling factors 5/2 and 7/3.
 
In situations where overlaps are suggestive but not conclusive, the standard approach is to ask if the ansatz wave function is a good ground state for a model interaction and then to establish adiabatic continuity along a line connecting the model interaction and the physical Coulomb interaction. No local interaction is known which produces our $\nu = 6/13$ wave function as the exact ground state. We instead draw inspiration from the Pfaffian wave function, which is the exact solution for a three-body interaction. It was previously shown that particle-hole symmetrization or a mean-field approximation of this interaction produces a two-body interaction whose ground sate is also very close to the Pfaffian state~\cite{Peterson08a,Sreejith17}. This interaction, denoted $H_2$, is defined by the pseudopotentials $V_1=3V_3$, and $V_m = 0$ for $m > 3$. The pseudopotential denoted by $V_{m}$ is the energy of a pair of electrons in a state with relative angular momentum $m$. We define an interaction $H(\lambda)=(1-\lambda)H_{\rm SLL}+\lambda H_2$, where $H_{\rm SLL}$ is the SLL Coulomb interaction and the value of $V_1$ in $H_2$ is taken to be that of the SLL Coulomb pseudopotential.

\begin{figure}[t]
\begin{center}
\includegraphics[width=0.48\textwidth,height=0.26\textwidth]{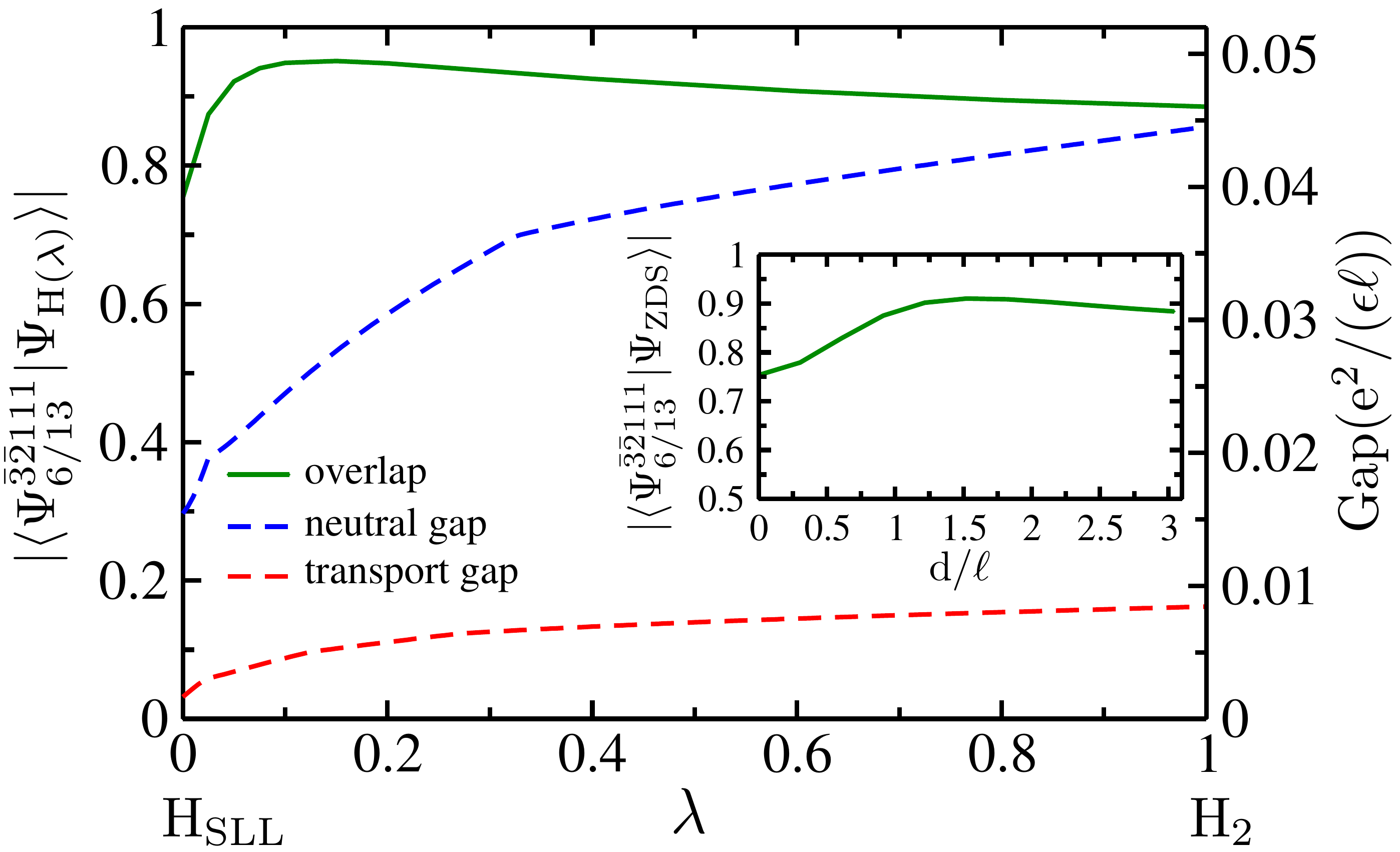}
\caption{The transport (red line) and neutral (blue line) gaps as the Hamiltonian is tuned from the second Landau level Coulomb interaction, modeled using truncated disk pseudopotentials, to the model Hamiltonian $H_{2}$, for $N=12$ electrons seeing a flux of $2Q=28$ in the spherical geometry. The model Hamiltonian $H_{2}$ is defined by the set of pseudopotentials $V_{1}=V^{\rm SLL}_{1}$ (same as the second Landau level Coulomb), $V_{3}=V_{1}/3$, and the rest $V_{m}=0$. The ``transport" gap for this system is defined as $[E(2Q=29)+E(2Q=27)-2E(2Q=28)]/6$, where $E(2Q)$ is the ground state energy at flux $2Q$ and the factor of 6 accounts for the fact that the removal or addition of one flux quantum produces 6 fundamental quasiparticles or quasiholes. The neutral gap for this system is defined as the difference between the two lowest energies at a fixed flux $2Q=28$. This figure also displays the overlap (green line) of the ground state with the $\bar{3}\bar{2}111$ parton state. Inset: Overlap between the parton state and the exact ground state of the Zhang-Das Sarma Hamiltonian $H_{\rm ZDS}$ as a function of thickness parameter $d/\ell$. }
\label{fig:overlap_gap_parton_6_13}
\end{center}
\end{figure}

In Fig.~\ref{fig:overlap_gap_parton_6_13} we show the overlap of the $\bar{3}\bar{2}111$ parton state with the exact ground state of $H(\lambda)$ for $0 \le \lambda \le 1$. We furthermore show the transport and neutral gaps for the latter. The high overlap near the second LL Coulomb point and the robust gaps support the assertion that the parton state is stabilized for a range of interaction potentials near the Coulomb interaction.

To assess the qualitative effect of the quantum well's finite width, we study the overlap of the $\bar{3}\bar{2}111$ ansatz with the numerically exact ground state for an interaction proposed by Zhang and Das Sarma (ZDS) that takes into account the finite quantum well width $d$~\cite{Zhang86}: $H_{\rm ZDS}=1/\sqrt{r^2+(d/2)^2}$. As shown in the inset of Fig.~\ref{fig:overlap_gap_parton_6_13}, the overlap in fact improves when the finite width is taken into account. This further underscores the robustness of the $\bar{3}\bar{2}111$ ansatz. LL mixing can also provide quantitative corrections, but a proper treatment of this effect is outside the scope of this work.

We believe that these comparisons make a clear case for the plausibility of the $\bar{3}\bar{2}111$ ansatz. In the remainder of the Letter we deduce the experimental consequences of our theory, which allow its validity to be assessed. 

The most immediate ramification of our proposal is the sequence $\nu=2n/(5n-2)$ arising from the $\bar{n}\bar{2}111$ parton states. The first three members of the sequence occur at $2+2/3$, $2+1/2$, and $2+6/13$, and have been observed. This provides a natural explanation for why $2+6/13$ is observed, which appears ``out of order" from the perspective of the LLL CF theory. A definitive observation of the next fraction $2+4/9$ or its hole partner will lend further credence to the parton paradigm for the second LL FQHE, although it is possible that this and further fractions are swamped by bubble phases.

The quasiparticles of the $\bar{3}\bar{2}111$ state obey Abelian braid statistics. An additional CF particle in the factor $\Phi_{\bar{3}}$ has charge $q_{\bar 3}=2e/13$, whereas that in the factor $\Phi_{\bar{2}}$ has a charge $q_{\bar 2}=3e/13$. A combination of a CF particle in $\Phi_{\bar{3}}$ and a CF hole in $\Phi_{\bar{2}}$ leads to the smallest charge, of magnitude $q_{\bar 2}-q_{\bar 3}=e/13$. At this stage, it has not been possible to reliably estimate the thermodynamic values of the gaps predicted by our parton ansatz. The transport gap of $0.0016 e^2/\epsilon \ell$ for $N=12$ particles, while much smaller than the gap of $0.1 e^2/\epsilon \ell$ at $\nu=1/3$, far exceeds the gap of $10.5~{\rm mK} \sim 0.0001 e^2/\epsilon \ell$ measured by Kumar {\em et al.}~\cite{Kumar10}. A significant discrepancy exists for the gaps of other FQHE states as well, especially for the more delicate ones, presumably arising from a combination of disorder, finite width effects, and LL mixing. We note that in the thermodynamic limit, we expect the neutral gap to be smaller than or equal to the transport gap; the large deviation between the two for 12 particles indicates strong finite size effects for excitations, as has been found for other fractions in the second LL \cite{Balram13b,Johri14,Morf02}.

To deduce other topological consequences, we consider the low-energy effective theory of the edge, which is described by the Lagrangian density~\cite{Wen91b,Wen92b,Moore98}: 
\be
\mathcal{L} = -\frac{1}{4\pi} K_{\rm IJ}\epsilon^{\mu\nu\lambda} a_{\mu}^{\rm I}\partial_{\nu} a_{\lambda}^{\rm J} - \frac{1}{2\pi} \epsilon^{\mu\nu\lambda}t_{\rm I}A_{\mu}\partial_{\nu} a_{\lambda}^{\rm I}.
\label{eq_eff_L}
\ee
Here we have used Einstein's summation convention, $\epsilon^{\mu\nu\lambda}$ is the completely antisymmetric Levi-Civita tensor, $A$ is the external electromagnetic vector potential, and $a$ denotes the internal gauge field. Naively, one might guess that there are a total of eight edge states: three from the factor $\Phi_{\bar{3}}$, two from $\Phi_{\bar{2}}$, and one from each factor $\Phi_1$. However, these are not all independent. Recalling that the density variations of all partons must be identified, which gives four constraints, one ends up with four independent edge states. The integer-valued symmetric $K$ matrix and the charge vector $t$ from Eq.~(\ref{eq_eff_L}) for the parton state are given by~\cite{SM}
\be
K =    \begin{pmatrix} 
      -2 & -1 & 0 & 1 \\
      -1 & -2 & 0 & 1 \\
       0 & 0 & -2 & 1 \\
       1 & 1 &  1 & 1 \\
   \end{pmatrix},\quad
t =    \begin{pmatrix} 
      \,0\, \\
      \,0\, \\
      \,0\, \\
      \,1\, \\
   \end{pmatrix}.   
\ee
The ground state degeneracy of the parton state on a manifold with genus $g$ is $|{\rm det}(K)|^{g}=13^{g}$. 

The $K$ matrix above has one positive and three negative eigenvalues; the $\bar{3}\bar{2}111$ state thus hosts one forward moving and three backward moving edge modes. The thermal Hall conductance $\kappa_{xy}$ takes a quantized value proportional to the chiral central charge $c$, which is the difference in the number of forward and backward moving modes: $\kappa_{xy} =c[\pi^2 k_{\rm B}^2 /(3h)]T$. For $\bar{3}\bar{2}111$ we thus predict a thermal Hall conductance of $\kappa_{xy} =-2[\pi^2 k_{\rm B}^2 /(3h)]T$. The Hall viscosity is also expected to be quantized~\cite{Read09}: $\eta_{\rm H} = \hbar \rho_{0} \mathcal{S}/4 =(-1/2)\hbar \rho_{0}$, where $\rho_{0}=\nu/(2\pi \ell^{2})$ is the density and $\mathcal{S}=-2$ is the shift. 

In contrast, the $K$ matrix of the $6/13$ CF state is given by the 6$\times$6 matrix $K_{ij}=2+\delta_{ij}$ and charge vector $t=(1,1,1,1,1,1)^{T}$. It is an Abelian state with quasiparticle charge $-e/13$ and degeneracy of $13^{g}$ on a manifold of genus $g$. In contrast to $\bar{3}\bar{2}111$, the CF state has six forward moving edge states and no upstream neutral modes (assuming the absence of edge reconstruction); its thermal Hall conductance $\kappa_{xy} =6[\pi^2 k_{\rm B}^2 /(3h)]T$; and its Hall viscosity $\eta_{\rm H} = 2\hbar \rho_{0}$, corresponding to shift ${\cal S}=8$. 

Shot noise experiments have been used to measure the presence of upstream modes~\cite{Bid10,Dolev11,Gross12,Inoue14} and also the quantized thermal Hall conductance~\cite{Banerjee17,Banerjee17b} at other filling fractions; these experiments can test the predictions of the parton theory and thus discriminate between the topological structures of the 6/13 states in the LLL and the SLL. In particular, including the contribution arising from the filled LLL, the thermal Hall conductance of the $\bar{3}\bar{2}111$ ansatz vanishes. This is dramatically different from what one would expect from the CF state, which has $\kappa_{xy} =8[\pi^2 k_{\rm B}^2 /(3h)]T$.

The Levin-Halperin state~\cite{Levin09a} at $\nu=2+6/13$ is also Abelian, occurs at shift ${\cal S}=-2$, and has a thermal Hall conductance of $\kappa_{xy}=-2[\pi^2 k_{\rm B}^2 /(3h)]T$. Therefore, it may be in the same topological phase as the $\bar{3}\bar{2}111$ state.  

To gain insight into what makes the $\bar{n}\bar{2}111$ parton states special, we consider other parton states. The fact that the CF states $n11\cdots$ and $\bar{n}11\cdots$ capture the most prominent states of the LLL suggests that placing parton species into $\nu = 1$ states builds good correlations. The simplest generalization thus is to have $n n_2 11\cdots$, with $|n_2|=2$, where $\cdots$ indicates that more 1's may be added. The states $\bar{n}21$ at $\nu=2n/(3n-2)$ and $\bar{n}\bar{2}1$ at $\nu=2n/(n-2)$ do not produce fractions in the filling factor range of our interest. The $n21$ parton states at $\nu=2n/(3n+2)=2/5, 1/2, 6/11, \cdots$ appear, \emph{a priori}, as plausible as the ones we considered above. However, this family does not provide an account of the SLL FQHE: the first two states, namely the 2/5 CF and the 221 parton states, have been ruled out for $\nu = 2+2/5$ and $\nu = 2+1/2$~\cite{Wojs09,Bonderson12}, respectively, and no FQHE has been seen at the third fraction in the sequence, $\nu = 2+6/11$, or its hole partner, $\nu = 2+5/11$. We note, however, that this is an energetic issue; the $n21$ parton states are conceptually well defined and can possibly be stabilized by some other interaction (see, e.g.,  Refs.~\cite{Wu17b,Bandyopadhyay18,Kim18} for the 221 parton state). For the SLL in GaAs, the sequence we propose appears to be the most plausible scenario.

In summary, we have proposed that the parton ansatz ``$\bar{n}\bar{2}111$'' naturally captures an important sequence of observed fractional quantum Hall states in the second Landau level, explaining, in particular, the unusual stability of $2+6/13$ FQHE. We have further suggested experimental quantities that can reveal the underlying parton character of the $2+6/13$ state and demonstrate it to be topologically distinct from the 6/13 state in the lowest Landau level. The parton construction can be readily generalized to multicomponent systems involving spin, valley, layer, or orbital degrees of freedom. The viability and properties of these states remain to be explored.

\begin{acknowledgments}
The Center for Quantum Devices is funded by the Danish National Research Foundation. This work was supported by the European Research Council (ERC) under the European Union Horizon 2020 Research and Innovation Programme, Grant Agreement No. 678862. A.C.B. and M.R. also thank the Villum Foundation for support. The work at Penn State was supported by the U. S. Department of Energy, Office of Basic Energy Sciences, under Grant no. DE-SC0005042. MB is supported by NSF CAREER (DMR-1753240) and JQI-PFC-UMD. Some of the numerical calculations were performed using the DiagHam package, for which we are grateful to its authors. Some portions of this research were conducted with Advanced CyberInfrastructure computational resources provided by The Institute for CyberScience at The Pennsylvania State University. 
\end{acknowledgments}

$^{\dagger}$ ACB and SM contributed equally to the manuscript.

\pagebreak
\begin{widetext}
\begin{center}
\textbf{\large Supplemental Material on ``Fractional Quantum Hall Effect at $\nu=2+6/13$: The Parton Paradigm for the Second Landau Level''}
\end{center}

\setcounter{figure}{0}
\setcounter{equation}{0}
\renewcommand\thefigure{S\arabic{figure}}
\renewcommand\thetable{S\arabic{table}}
\renewcommand\theequation{S\arabic{equation}}

\section{Derivation of the low-energy effective theory of the parton 6/13 edge}
\label{appendix:eff_edge}
 The unprojected wave function of the parton sequence described in the main text can be re-written as:
\begin{equation}
\Psi^{\bar{n}\bar{2}111}_{2n/(5n-2)} = [\Phi_{n}]^{*}[\Phi_{2}]^{*}\Psi_{1/3}.
\end{equation}
Viewed in this way, we may express the state in terms of partons $\wp = f_{1}f_{2}f_{3}$, where the fermionic partons are in the following mean-field states: $f_{1}$ is in a $\nu=-n$ integer quantum Hall (IQH) state, $f_{2}$ is in a $\nu=-2$ IQH state and $f_{3}$ is in a $\nu=1/3$ Laughlin state (which in turn can be viewed as a product of three partons each residing in a $\nu=1$ IQH state). Following the discussions in the main text, the charges of these various partons are: $q_{1}=2e/(5n-2)$, $q_{2}=ne/(5n-2)$ and $q_{3}=-e-q_{1}-q_{2}=-6ne/(5n-2)$, where $-e$ is the electron charge (henceforth we set $e=1$ for convenience). For $n\neq 2$ the residual gauge symmetry of this parton ansatz is $U(1)\times U(1)$, associated with the transformations:
\begin{equation}
f_{1}\rightarrow e^{i\theta_{1}} f_{1},~f_{2}\rightarrow e^{-i\theta_{1}+i\theta_{2}} f_{2},~
f_{3}\rightarrow e^{-i\theta_{2}} f_{3}.
\end{equation}
Therefore, we have two internal $U(1)$ gauge fields which we shall denote by $h_{\mu}$ and $g_{\mu}$. The low-energy effective field theory for this parton mean-field state is given by the Lagrangian density:
\begin{eqnarray}
\mathcal{L}&=&\frac{1}{4\pi} \sum_{i=1}^{n}\alpha^{(i)}\partial \alpha^{(i)}+\frac{1}{2\pi}\sum_{i=1}^{n}(h+q_{1}A)\partial \alpha^{(i)}  \nonumber \\
&+& \frac{1}{4\pi} \sum_{j=1}^{2}\beta^{(j)}\partial \beta^{(j)}+\frac{1}{2\pi}\sum_{j=1}^{2}(g-h+q_{2}A)\partial \beta^{(j)}  \nonumber \\
&-& \frac{3}{4\pi} \gamma \partial \gamma+\frac{1}{2\pi}(-g+q_{3}A)\partial \gamma,
\label{eq:Lagrangian_density}
\end{eqnarray}
where we use the short-hand notation $\alpha\partial \alpha$ to denote $\epsilon^{\mu\nu\lambda}\alpha_{\mu}\partial_{\nu}\alpha_{\lambda}$, where $\epsilon^{\mu\nu\lambda}$ is the completely anti-symmetric Levi-Civita tensor and Einstein's summation convention is assumed. In the above Lagrangian density, $A$ is the external electromagnetic vector potential, and $\alpha^{(i)}$, $\beta^{(j)}$ and $\gamma$ are $U(1)$ gauge fields describing the current fluctuations of the IQH states. Thus we have a $U(1)^{n+5}$ Chern-Simons theory ($n$ from $[\Phi_{n}]^{*}$, $2$ from $[\Phi_{2}]^{*}$ and $3$ from $\Psi_{1/3}$ adding up to $n+5$) which can be described by an integer valued symmetric $(n+5)\times (n+5)$ $K$ matrix. Since we have further combined the three gauge fields of the Laughlin state into one, we end up with an $(n+3)\times (n+3)$ $K$ matrix. \\

This theory can be simplified by integrating out the internal gauge fields $h$ and $g$. Integrating out $h$ results in the constraint:
\begin{equation}
\epsilon^{\mu\nu\lambda} \sum_{i=1}^{n} \partial_\nu \alpha_\lambda^{(i)} = \epsilon^{\mu\nu\lambda} \sum_{j=1}^{2} \partial_\nu \beta_\lambda^{(j)} .
\end{equation}
This can be solved by setting
\begin{align}
\sum_{i=1}^{n} \alpha^{(i)} = \sum_{j=1}^{2} \beta^{(j)} + c ,
\label{eq:constraint1}
\end{align}
where $c$ is a $U(1)$ gauge field that satisfies 
\begin{align}
\epsilon^{\mu\nu\lambda} \partial_\nu c_\lambda = 0. 
\end{align}
Furthermore, integrating out $g$ results in:
\begin{equation}
\epsilon^{\mu\nu\lambda} \partial_\nu \gamma_\lambda = \epsilon^{\mu\nu\lambda} \sum_{j=1}^{2} \partial_\nu \beta_\lambda^{(j)} ,
\end{equation}
which can be solved by setting
\begin{equation}
\gamma = \sum_{j=1}^{2} \beta^{(j)} + d,
\label{eq:constraint2}
\end{equation}
where similarly $d$ is a $U(1)$ gauge field that satisfies
\begin{align}
\epsilon^{\mu\nu\lambda} \partial_\nu d_\lambda = 0. 
\end{align}

When substituting Eq.~(\ref{eq:constraint1}) and (\ref{eq:constraint2}) into Eq.~(\ref{eq:Lagrangian_density}), all terms involving $c$ and $d$ vanish (some terms require integration by parts to see this), yielding a simplified 
$U(1)^{n+1}$ Chern-Simons theory which can be described by an integer valued symmetric $(n+1)\times (n+1)$ $K$ matrix.

\subsection{Specializing to the $n=3$, $\nu=6/13$ state}

Let us now focus our attention to the $n=3$ case. Following Eqs.~(\ref{eq:constraint1}) and (\ref{eq:constraint2}) we have:
\begin{equation}
\alpha^{(3)}=\gamma - \alpha^{(1)} - \alpha^{(2)} -d - c,~
\beta^{(2)} =\gamma - \beta^{(1)} - d.
\end{equation}
Substituting this back into the Lagrangian density given in Eq.~(\ref{eq:Lagrangian_density}), and using the fact that terms involving $c$ and $d$ vanish,
we obtain:
{\small
\begin{eqnarray}
\mathcal{L}&=&\frac{1}{4\pi}\left( \alpha^{(1)}\partial \alpha^{(1)}+\alpha^{(2)}\partial \alpha^{(2)}+
(\gamma - \alpha^{(1)} -\alpha^{(2)}) \partial (\gamma - \alpha^{(1)} -\alpha^{(2)}) \right)  \nonumber \\
&+& \frac{1}{4\pi}\left( \beta^{(1)}\partial \beta^{(1)} + (\gamma-\beta^{(1)})\partial (\gamma-\beta^{(1)})\right) \\
&-& \frac{3}{4\pi} \gamma \partial \gamma + \frac{1}{2\pi}A\partial \gamma. \nonumber 
\end{eqnarray}
}
Let us now define a new set of gauge fields:
\begin{equation}
 (a^{1},a^{2},a^{3},a^{4}) = (\alpha^{(1)},\alpha^{(2)},\beta^{(1)},\gamma),
\end{equation}
such that we can write the Lagrangian in the familiar form:
\begin{equation}
 \mathcal{L} = -\frac{1}{4\pi} K_{\rm IJ}a^{\rm I}\partial a^{\rm J} + \frac{1}{2\pi} t^{\rm I}A\partial a^{\rm I}.
\end{equation}
Here the charge vector is $t=(0,0,0,1)^{\rm T}$ and the $K$ matrix given by
\begin{equation}
K =   
 \begin{pmatrix} 
      -1 & 0 & 0 & 0 \\
       0 &-1 & 0 & 0 \\
       0 & 0 &-1 & 0 \\
       0 & 0 & 0 & 3 \\
   \end{pmatrix} + 
 \begin{pmatrix} 
       0 & 0 & 0 & 0 \\
       0 & 0 & 0 & 0 \\
       0 & 0 &-1 & 1 \\
       0 & 0 & 1 &-1 \\
   \end{pmatrix} + 
 \begin{pmatrix} 
      -1 &-1 & 0 & 1 \\
      -1 &-1 & 0 & 1 \\
       0 & 0 & 0 & 0 \\
       1 & 1 & 0 &-1 \\
   \end{pmatrix}   
      =
\begin{pmatrix} 
      -2 & -1 & 0 & 1 \\
      -1 & -2 & 0 & 1 \\
       0 & 0 & -2 & 1 \\
       1 & 1 &  1 & 1 \\
   \end{pmatrix}. 
\end{equation}
The filling fraction is:
\begin{equation}
\nu =  t^{\rm T}\cdot K^{-1} \cdot t=K^{-1}_{44} = 6/13,
\end{equation}
as expected. The ground state degeneracy on a manifold with genus $g$ is:
\begin{equation}
 \text{ground state degeneracy} = |{\rm Det}(K)|^{g} =13^{g}.
\end{equation}
This $K$ matrix has one positive and three negative eigenvalues which indicates that the $\bar{3}\bar{2}111$ state hosts one forward moving and three backward moving edge modes. The statistics and the fractional charges of the quasiparticles can be read off from the $K$ matrix in the usual way. 

\subsection{Shift and coupling to curvature}
We can also include coupling to curvature in order to compute the shift on a sphere from the topological field theory. Recall that for an IQH state filling the $n^{\rm th}$ Landau level, the effective Lagrangian density including the coupling to curvature is:
\begin{equation}
\mathcal{L} = -\frac{1}{4\pi} \alpha \partial \alpha + \frac{1}{2\pi} s \omega \partial \alpha + \frac{1}{2\pi} A \partial \alpha,
\end{equation}
where the spin is $s=(n-1/2)$ and $\omega$ is the spin connection. Therefore when we include the coupling to curvature in the above Lagrangian we get an additional term:
\begin{equation}
\delta\mathcal{L} = -\frac{1}{2\pi} \sum_{i=1}^{n} (i-1/2)\omega \partial \alpha^{(i)} 
-\frac{1}{2\pi} \sum_{j=1}^{2} (j-1/2)\omega \partial \beta^{(j)} 
+ \frac{1}{2\pi} \frac{3}{2} \omega \partial \gamma. 
\end{equation}
Specializing again to the case of $n=3$ and following a similar line of argument as above we end up with the following additional term in Lagrangian density which describes coupling to the curvature:
\begin{eqnarray}
\delta\mathcal{L} &=& -\frac{1}{2\pi} \left( (1/2)\omega \partial \alpha^{(1)} + (3/2)\omega \partial \alpha^{(2)}+
(5/2) \omega \partial (\gamma - \alpha^{(1)} -\alpha^{(2)})\right) \nonumber \\
&-&\frac{1}{2\pi} \left( (1/2)\omega \partial \beta^{(1)} + (3/2)\omega \partial (\gamma - \beta^{(1)})\right)+ \frac{1}{2\pi} \frac{3}{2} \omega \partial \gamma \nonumber \\
&=& \frac{1}{2\pi} \left( 2\omega \partial \alpha^{(1)} + \omega \partial \alpha^{(2)} + \omega \partial \beta^{(1)}-(5/2)\omega \partial \gamma \right) =\frac{1}{2\pi} \mathfrak{s}^{\rm I}\omega\partial a^{\rm I}, 
\end{eqnarray}
where we have defined the spin vector $\mathfrak{s}=(2,1,1,-(5/2))^{\rm T}$. The shift on the sphere is therefore:
\begin{equation}
 \mathcal{S}=\frac{2}{\nu} t^{\rm T}\cdot K^{-1} \cdot \mathfrak{s}=-2,
\end{equation}
which is consistent with the value obtained in the main text from the wave function.

\end{widetext}

\bibliography{../biblio_fqhe}

\begin{thebibliography}{72}
\bibitem[{foo()}]{footnote:two_thirds_variants}
\bibinfo{note}{The choice $n=1$ in Eq.~(\ref{eq:parton_family}) produces a
  state that differs from the LLL $\nu = 2/3$ state $[\Phi_2]^*\Phi_1^2$ only
  through the additional factor $|\Phi_1|^2$. It was shown in
  Ref.~\cite{Balram16b} that $\bar{2}\bar{1}111$ and $\bar{2}11$ are very
  nearly the same after LLL projection.}
\end{thebibliography}
\bibliographystyle{apsrev_nourl}
\end{document}